\documentclass{ws-procs961x669}            % default, citations in superscript
\begin{document}
\title{The curious case of Betelgeuse}

\author{Jacco Th. van Loon}

\address{Lennard-Jones Laboratories,\\
Keele University, ST5 5BG, UK\\
E-mail: j.t.van.loon@keele.ac.uk\\
www.astro.keele.ac.uk/$\sim$jacco}

\begin{abstract}
Betelgeuse is the nearest red supergiant, one of the brightest stars in our
sky, and statistically speaking it would be expected to be "typical". Yet it
exhibits many features that seem "curious", to say the least. For instance it
has a high proper motion. It rotates fast. It has little dust. It dimmed
unexpectedly. Is any of these, and other, phenomena atypical, and taken
together does it make Betelgeuse atypical? This is important to know, because
we need to know whether Betelgeuse might be a prototype of red supergiants in
general, or certain subclasses of red supergiants, since we can study it in
such great detail. It is also important to know as it may be a link to
understanding other, apparently atypical cases such as supernova 1987A, and
maybe even such exotica as Thorne-\.Zytkov objects. Studying this question in
itself helps us understand how we deal with rarity and coincidence in
understanding the Universe we live in.
\end{abstract}

\keywords{Red supergiants -- stellar variability -- dust formation --
          supernova progenitors -- Philosophy of Science -- Betelgeuse.}

\bodymatter

\section{Prologue}

Not far in the sky from the most famous supernova remnant, the Crab Nebula,
the most famous red supergiant, Betelgeuse, has been a familiar sight for
peoples in either hemisphere. Distinctively orange, and one of the brightest
lights on the nightly celestial firmament, its prominent position in the
corner of arguably the most iconic constellation, Orion, Betelgeuse commands
our attention and this in itself makes it special. This perspective aims to
challenge the uniqueness and surprising behaviour of many a favourite star,
and what it tells us about us. It could be seen as a more philosophical sequel
to the Paris 2012 workshop on Betelgeuse\cite{vanLoon13}.

\section{Is Betelgeuse typical?}

It's got to be, right? Surely, as the nearest example of a red supergiant,
Betelgeuse must be ordinary? This is because the chances are favourable for a
draw of one from among many to be drawn from the most common.

While this may often be true, and hence unremarkable and thus unnoticed, it is
not always the case, as is expected when considering many draws from samples
that do include rare examples. (The basis for many false associations, let
alone causal relationships.) For instance, the first supernova witnessed by
naked eye for three centuries, and the best studied, supernova 1987A wasn't!
It had been expected to have been a star like Betelgeuse to have exploded, and
yet it was all but. (Though they may have much more in common than at first
sight\cite{vanLoon13}.)

We are not placed in a typical red supergiant environment. While within a
Bubble created by multiple supernov{\ae}, we are not currently situated in a
spiral arm. So perhaps Betelgeuse is typical for this particular environment,
but not necessarily for red supergiants that are largely found in the denser,
more actively star-forming regions of a spiral galaxy such as the spectacular
H\,{\sc ii} regions in the prominent spiral arms several times more distant,
or the even more distant massive clusters near the Galactic Centre.

But all things considered, Betelgeuse does not stand out in either luminosity,
temperature, mass-loss rate or dust content -- in fact it looks like a fairly
commonly identified SN\,II-P progenitor!\cite{Smartt15} (But this may partly
be a selection bias of the progenitor identification strategies.)

\section{Is Betelgeuse atypical?}

Just like most people look similar -- four limbs, a head with a pair of eyes
and ears and a nose upfront -- when inspected more closely individual traits
set us apart. This doesn't make each specimen a species, and in the end, we
have much more in common than differentiates us, both in terms of the way we
look and the way we behave.

Among Betelgeuse's perplexities count its location at some distance from sites
of recent star formation, its large space motion, and a picture-perfect bow
shock\cite{Decin12}. Upon close inspection, it spins at a baffling
rate\cite{Uitenbroek98,Kervella18} and also is
super-nitrogenous\cite{Lambert84}.

And then came the `Great Dimming'\cite{Montarges21}. (Which may have nothing
or all to do with the above.) A drop in brightness by half relegated
Betelgeuse to the realm of much more inconspicuous stars, but sent alarm bells
ringing with fears -- or excitement -- for its imminent luminous demise.

But none have been scrutinised as much as Betelgeuse. Brief late phases of
under-represented stars at birth, red supergiants themselves are rare, so few
are seen up as close as Betelgeuse. It is not how well we know Betelgeuse, but
how poorly we know the other red supergiants that defines its curiousness.

\section{Our one-sided view of Betelgeuse}

Because of the attention it has recently received, we shall first unpick how
rare is Betelgeuse's Great Dimming, before we return to its full portfolio of
properties.

By our very own nature as observers and thinking minds, and our position in
the Universe, we suffer from anthropocentric bias: just because {\em we}
didn't see it before, or elsewhere, doesn't make it rare per se. The
$20^{\rm th}$-century artist Pablo Picasso exposed this imperfect picture by
recovering what is hidden from view in a form referred to as ``cubism''. But
how do we do that in reality? Do we know it? Do we infer it? Or do we ignore
it? If you look out into a field and see what looks like a black lamb, what do
you conclude? That:
\begin{itemize}
\item lambs are black?
\item lambs exist that are black?
\item lambs exist of which at least one side is black?
\item apparitions exist that look like lambs of which at least one side is
black?
\end{itemize}
Despite well-known examples of non-isotropy in astronomy (cf.\ exoplanet
transits, $\gamma$-ray bursts, pulsars) it is too often overlooked.

Betelgeuse forms dust clouds episodically, as convection cells cause dark
spots\cite{Haubois09} that induce cooler circumstellar conditions where dust
can form\cite{Kervella11}, likely in an already cooler parcel of lifted
gas\cite{Kervella09}. In the case of Betelgeuse this is exacerbated because
Betelgeuse is on the cusp between a warm chromosphere and dusty
wind\cite{vanLoon05}. Counting the number of dust clouds and considering their
projected distances from Betelgeuse, a dust cloud forms about once every five
years!

Imagine Betelgeuse were cubic and we only faced one side at a time (compare a
standard die), then we'd only see one in every six events randomly happening
on one of its sides. (In reality the polar and equatorial regions may display
different behaviour especially in the case of a rapidly spinning Betelgeuse.)
This means we {\em may} have had a Great Dimming {\em on our side} once every
few decades (cf.\ the visual lightcurve\cite{Joyce20}), but we {\em certainly}
missed Great Dimmings happening {\em on another side}!

It is tempting to expect a dust cloud to form at every minimum in the
$\sim2000$ day cycle. But that depends on whether this periodicity represents
radial pulsation (in which case it is true) or convectional modulation -- in
which case the five-yearly cloud would always be in front, though other clouds
would be forming more frequently throughout, above cool convection cells not
seen by us, hence being in tension with observational evidence. It is more
likely that a dust cloud forms at a pulsational minimum but only above a cool
convection cell (the latter causing the $\sim400$-day cycle, but again that's
our biased perspective, there could be a cool convection cell {\em somewhere}
on the surface at all times). In fact, forming at some height above the
surface, such dust cloud is less likely to be seen directly in front of the
star than the convection cell is -- at one stellar radius above the surface
this chance is already diminished by a factor four. (Note that even the cloud
purported to have caused the Great Dimming did not cover the entire face of
Betelgeuse\cite{Montarges21}.) The Great Dimming might have been our treat,
but bread and butter for Betelgeuse.

Previous dimmings have been disputed, but it is not outrageous if of a few
expected {\em chance} occurrences (as opposed to {\em predicted} events) just
one has materialised. A century is not a very long time span in the life of a
star, even that of a red supergiant (one pro mille!), and Betelgeuse may have
exhibited many more dimmings, and may exhibit many more to come. Our Sun has
arguably been more dramatic, as we know it so well: the sunspot cycle is a
fairly regular rythm of activity, but we are still astounded by the Maunder
Minimum of the late-$17^{\rm th}$ and early-$18^{\rm th}$ centuries -- what if
the telescope had only been invented a century later? (We would never have
known.)

\section{How common are anomalies?}

If Betelgeuse ought to be common for being the nearest red supergiant, then so
should be Antares for being only the second-nearest red supergiant. But as we
argued, this is not a given. It would be much less likely, though, if both
Betelgeuse {\em and} Antares were special. This then begs the question: is
Betelgeuse different from Antares?

Antares has similar luminosity, temperature, convection\cite{Ohnaka17}, mass
loss, (low) dust content and lightcurve (with similar dimmings). Like
Betelgeuse, it produces discrete dust clouds\cite{Ohnaka14,Cannon21}. But
Antares is a {\em slow} rotator and is accompanied by a hot star (on a
sufficiently wide orbit not to be directly affected by it). How different does
that make them?

Imagine stars are characterised by five features, that each have a 1:10 chance
of being anomalous (imagine a ten-sided die). Then 1:2 stars are expected to
be anomalous. Every other star. So a star is just as likely to be common as it
is to be anomalous, despite the majority of stars to be common in any given
feature. It should not come as a surprise if both Betelgeuse and Antares had
some properties that are rare among the general population of red supergiants.
This is what makes individual people distinctive.

Imagine another scenario, where a penguin among other penguins in a colony is
distinguished by:
\begin{itemize}
\item the colour of its coat: blue (as opposed to brown).
\item the texture of its coat: smooth (as opposed to fluffy).
\item the colour of its cheeks: yellow.
\item the colour of the underbill: orange.
\end{itemize}
Does that make this individual penguin anomalous in four ways? Most definitely
not! It is anomalous in just one way: being an adult (surrounded by infants)!

\begin{figure}[h]
\begin{center}
\includegraphics[width=3in]{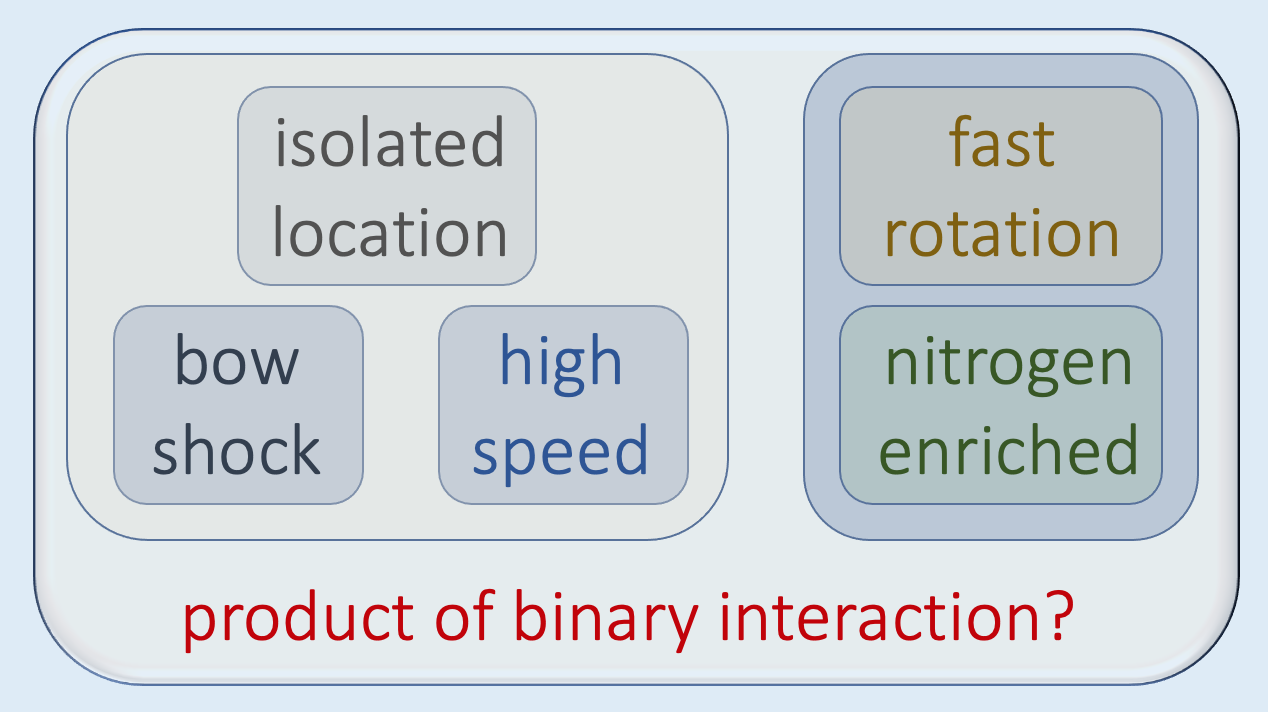}
\end{center}
\caption{Betelgeuse has various characteristics which have been treated as
peculiarities. Irrespective of whether these are indeed oddities at all, they
can be reduced to fewer characteristics -- possibly just one -- for which
there may well be a totally reasonable -- if not banal -- explanation.}
\label{vanLoon:fig1}
\end{figure}

Returning to the curious case of Betelgeuse, its isolated location, high speed
and bow shock are all related, and so may be its fast rotation and nitrogen
enrichment -- both can be reconciled into one: binary interaction
(Fig.~\ref{vanLoon:fig1}). And a common one at that: 10--40 per cent of
massive stars will be affected by a companion\cite{Sana12,deMink14,Eldridge17}.

Then which is more special: Betelgeuse or Antares? Neither -- both cases are
expected equally: the result of binary interaction (Betelgeuse) and multiple
but unaffected (Antares). The nearby Universe thus seems to be living up to
expectations remarkably well!

\section{How to find out how common is Betelgeuse?}

Moving onwards, if we want to find out what red supergiants typically look
like, how they generally evolve, and how and why some may deviate from the
norm, we will not succeed in this by studying Betelgeuse harder and longer,
but by statistical studies of samples of other red supergiants, such as:

\begin{itemize}
\item lightcurve studies based on sparse and/or time-limited data but for
large samples in nearby galaxies (`Great Dimmings' may be seen in other
galaxies, too\cite{Jencson21}).
\item nitrogen abundance measurements and their link with rotation.
\item radial velocity monitoring and spectral energy distributions, to
determine binary fractions.
\item three-dimensional space motion studies in conjunction with rejuvenation
scenarios, to determine past binary interactions.
\item interferometric size measurements in relation to the location on the
Hertzsprung--Russell Diagram, and mass (loss) measurements from seismology.
\end{itemize}

\begin{figure}[h]
\begin{center}
\includegraphics[width=3in]{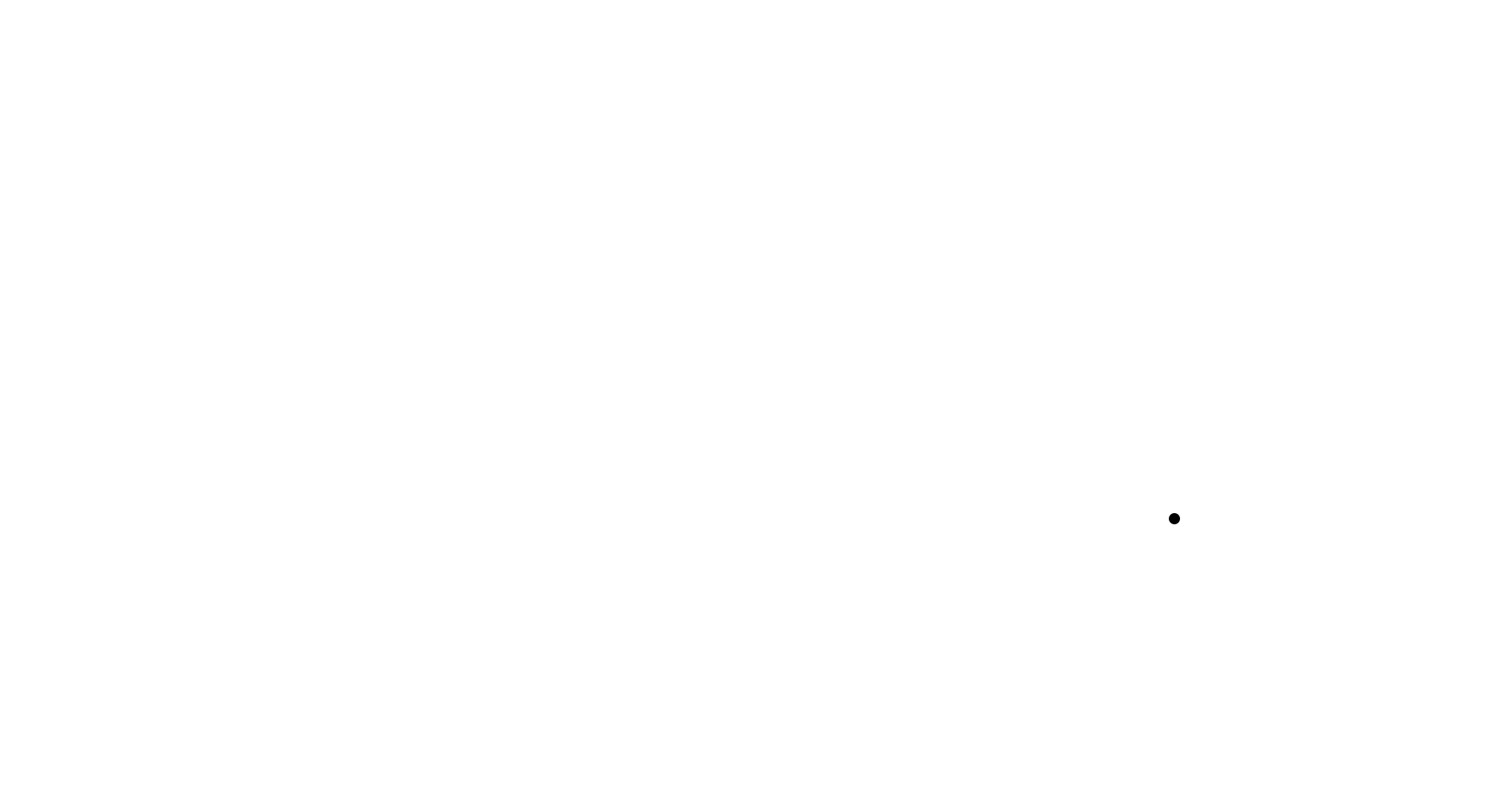}
\end{center}
\caption{What draws your eye? The white canvas that occupies most of what
you're looking at? Or the black dot offset from the centre, which only covers
a tiny fraction of your view? We are drawn to the exceptional, not the norm.}
\label{vanLoon:fig2}
\end{figure}

\section{Epilogue}

The take-away message from this discussion is that when making inferences we
need to account for biases. We tend to look for the distinctive, not the
common. For instance, when we look at a white panel with a black dot
(Fig.~\ref{vanLoon:fig2}) our attention is drawn to the dot even though it
occupies much less space than the rest of the canvass. Just as we occupy
ourselves with the few per cent of baryonic Universe, preferentially in its
condensed forms, leaving the vastness of space largely neglected. If we saw
less of the canvass, we might have missed the black dot altogether, but if we
saw more we might find out black dots to be rather common!

\section*{Acknowledgements}
I'd like to thank Costantino Sigismondi for organising the Betelgeuse session
at the $16^{\rm th}$ Marcel Grossmann meeting and for inviting me to speak.

\end{document}